\documentclass[11pt]{article}

\usepackage[final]{acl}

\usepackage{times}
\usepackage{latexsym}

\usepackage[T1]{fontenc}

\usepackage[utf8]{inputenc}

\usepackage{microtype}

\usepackage{inconsolata}

\usepackage{graphicx}

\usepackage{algorithm}
\usepackage{algorithmic}
\usepackage{makecell}
\usepackage{newfloat}
\usepackage{listings}
\usepackage{times}
\usepackage{epsfig}
\usepackage{float}
\usepackage{placeins}
\usepackage{enumitem}
\usepackage{tabularx}
\usepackage{xstring}
\usepackage{xspace}
\usepackage[hang,flushmargin]{footmisc}
\usepackage[utf8]{inputenc} 
\usepackage[T1]{fontenc}    
\usepackage{url}            
\usepackage{booktabs}       
\usepackage{amsfonts}       
\usepackage{nicefrac}       
\usepackage{xcolor}         
\usepackage{amsmath, amssymb, overpic, textpos}
\usepackage{graphicx}
\usepackage{multirow}
\usepackage{color, colortbl}
\usepackage{tabulary}
\usepackage{listings}
\usepackage{multicol}
\usepackage{xspace}
\usepackage{graphbox}
\usepackage{arydshln}
\usepackage{adjustbox}
\usepackage{kotex}
\usepackage{caption}
\usepackage{subcaption}
\usepackage{makecell}
\usepackage{tabularray}
\usepackage{bm}

\definecolor{Green}{rgb}{0.2, 0.7, 0.1}
\definecolor{Gray}{HTML}{B0B0B0} 
\definecolor{Blue}{HTML}{4A90E2} 
\definecolor{Yellow}{HTML}{FFA700} 
\definecolor{prompt_blue}{HTML}{1f78b4}
\definecolor{prompt_red}{HTML}{d45c43}
\definecolor{plus}{HTML}{0071bc}
\definecolor{minus}{RGB}{153,10,10}
\definecolor{SecondBest}{HTML}{E0F0FA}
\definecolor{Best}{HTML}{BAD8F2}

\usepackage[table]{xcolor}

\usepackage[utf8]{inputenc}

\usepackage{tcolorbox}
\usepackage{etoc}
\usepackage{titletoc}

\title{SubT: Subspace Tuning for Few-shot Generalization\\of Audio-Language Models}

\author{
    Jaehyuk Jang \qquad
    Kangwook Ko \qquad
    Wonjun Lee \qquad
    Changick Kim\\[0.5em]
Korea Advanced Institute of Science and Technology, South Korea \\[0.5em]
    {\tt\small \{jhyuk, kw.ko, dpenguin, changick\}@kaist.ac.kr}\\
}

\begin{document}

\maketitle


\begin{abstract}
Few-shot parameter-efficient adaptation of pretrained Audio--Language Models (ALMs) often improves seen-class performance at the cost of unseen-class generalization, leading to the base-to-new trade-off. We study this failure through zero-shot drift in the text embedding space: few-shot tuning can distort inter-class structure and move adapted embeddings away from their pretrained anchors. We therefore propose \textbf{Subspace Tuning (SubT)}, an embedding-space adaptation method that combines a geometry-aware shared transformation with anchoring to the zero-shot prototypes. The learned transformation is transferred to unseen classes, with subspace-aware gating to mitigate negative transfer. Across 11 audio benchmarks, SubT achieves strong few-shot generalization while operating directly on precomputed text embeddings without text-encoder backpropagation. Code is available at \href{https://github.com/jhyukjang/SubT}{https://github.com/jhyukjang/SubT}.

\end{abstract}


\section{Introduction}
\label{sec:intro}

Pretrained Audio--Language Models (ALMs)~\citep{laionclap,wav2clip,clap22,clap23} have enabled strong zero-shot transfer across diverse audio tasks, including sound event and acoustic scene classification~\citep{audioset}, emotion recognition~\citep{cmu-mosei, meld}, and audio captioning~\citep{audiocaps}, by aligning audio and text in a shared embedding space.
In practice, however, downstream deployment often requires few-shot adaptation under dataset- and domain-specific shifts.
Therefore, parameter-efficient adaptation methods, such as prompt tuning (e.g., CoOp~\citep{coop}) and adapters (e.g., CLIP-Adapter~\citep{clipadapter}), have become a standard choice.

\begin{figure}[t!]
  \centering
  \includegraphics[width=\linewidth]{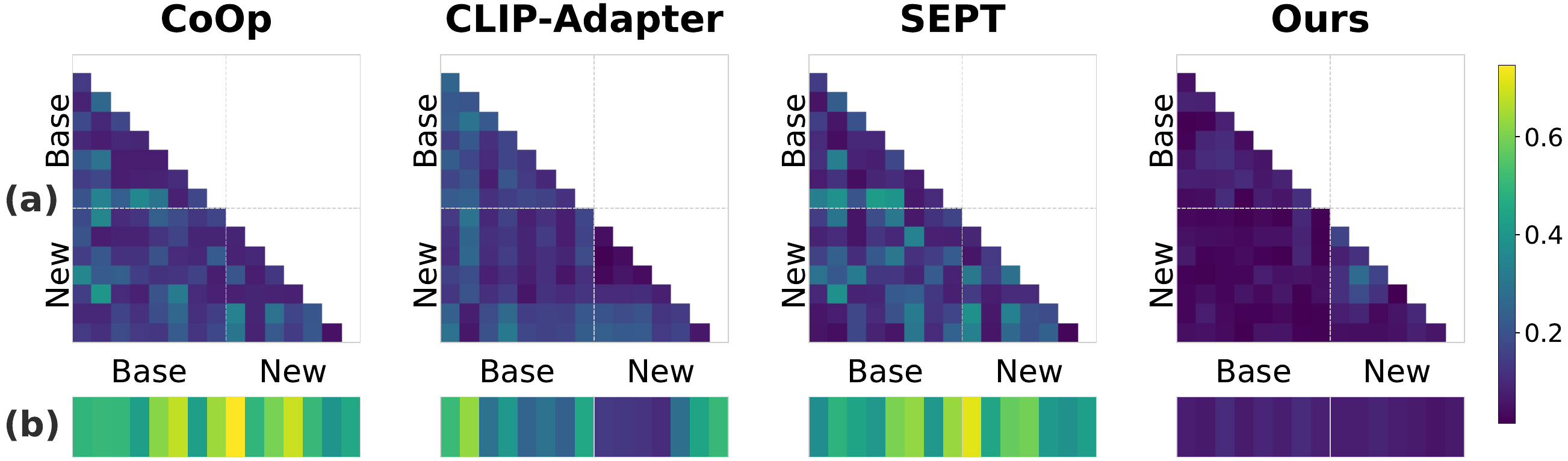}
    \caption{Comparison of few-shot adaptation methods against the zero-shot reference on TUT2017 in terms of (a) Gram drift in the class-wise cosine-similarity matrix and (b) Magnitude drift from the zero-shot prototypes. Compared with prior few-shot adaptation methods, our method more effectively controls both forms of  drift.}    
  \label{fig:motivation}
\end{figure}

\begin{figure*}[t!]
  \centering
  \includegraphics[width=.98\textwidth]{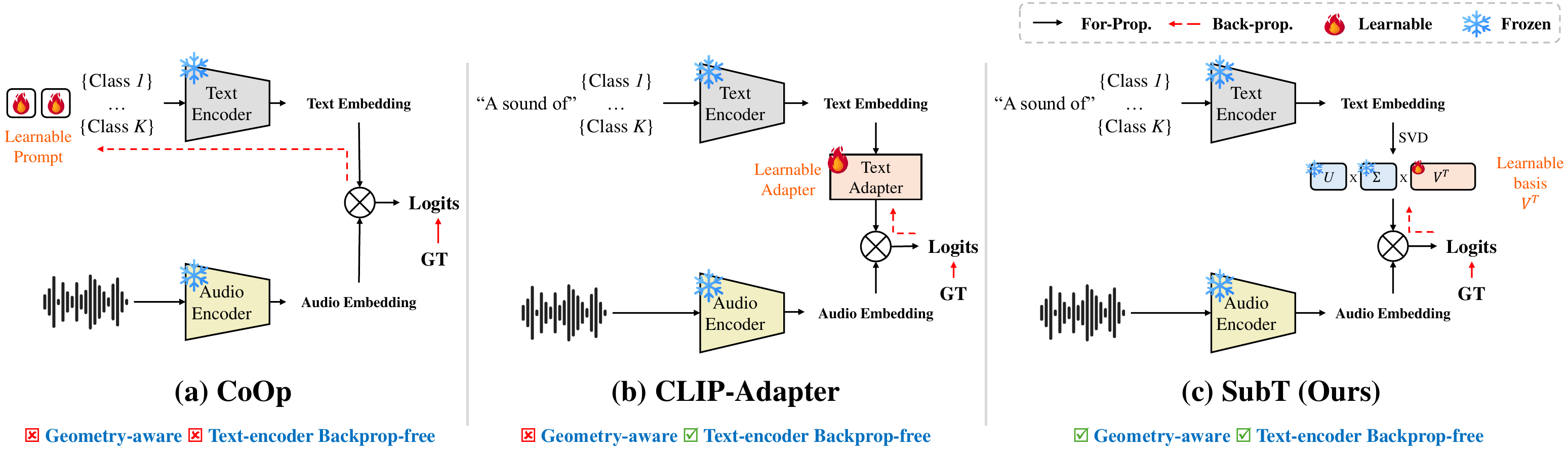}
    \caption{Architectural comparison of parameter-efficient adaptation methods. (a) Prompt tuning updates the input side of the text encoder and requires text-encoder backpropagation, while (b) CLIP-Adapter performs output-space adaptation without explicitly incorporating the zero-shot class geometry. In contrast, (c) our Subspace Tuning (SubT) operates directly on text embeddings and parameterizes adaptation through a geometry-aware shared basis, which can subsequently be transferred to unseen classes through the base semantic subspace.}
  \label{fig:method_comparison}
\end{figure*}

However, these methods often improve performance on seen \emph{base} classes at the expense of unseen \emph{new} classes, leading to the well-known base-to-new trade-off.
To alleviate this problem, prior work introduces additional inductive biases, including architectural enhancements~\citep{cocoop,dept,maple}, anchor-based regularization~\citep{kgcoop,lasp}, gradient projection~\citep{prograd}, and topology-preserving regularization~\citep{sept}.

Despite these efforts, drift in the pretrained text space remains insufficiently controlled.
To better understand this limitation, we examine how few-shot adaptation perturbs pretrained class text embeddings relative to the zero-shot reference.
Let $F_0,F \in \mathbb{R}^{N \times D}$ denote the row-wise normalized zero-shot and adapted class text embeddings, respectively.
We further define $G_0 = F_0 F_0^\top$ and $G = F F^\top$ as the corresponding Gram matrices, i.e., the pairwise cosine-similarity structures among classes.
As shown in Fig.~\ref{fig:motivation}, existing adaptation methods exhibit drift from the zero-shot reference in two complementary ways:
(a) distortion of the inter-class relational structure, measured by the element-wise absolute difference ($|G-G_0|$), and
(b) displacement of the adapted embeddings from their zero-shot anchors, measured by the cosine distance ($1-\cos(F,F_0)$).
Such drift can weaken the pretrained semantic reference that supports transfer to unseen classes, particularly when few-shot supervision drives adaptation toward the observed base classes.

Motivated by this observation, we propose \textbf{Subspace Tuning (SubT)}, a geometry-aware embedding-space adaptation framework for base-to-new generalization.
Structured Subspace Parameterization reparameterizes base-class adaptation through a shared basis factor defined by the zero-shot SVD coordinates, parameterizing optimization according to the pretrained class geometry rather than directly optimizing class prototypes.
Residual Anchoring complements this parameterization by retaining an explicit connection to the zero-shot prototypes and limiting excessive displacement under scarce supervision.
At inference time, the learned basis transformation is transferred to unseen classes through the base semantic subspace, while Subspace-aware Gating attenuates potentially harmful transfer for weakly aligned classes.

Extensive experiments on 11 audio datasets for base-to-new generalization and cross-dataset evaluation demonstrate strong few-shot generalization with SubT.
As shown in Fig.~\ref{fig:method_comparison}, SubT performs geometry-aware adaptation directly in the text embedding space while retaining the zero-shot semantic reference.
The resulting adaptation can be transferred to unseen classes without introducing new-class-specific parameters or backpropagating through the text encoder.


\section{Related Work}
\label{sec:related_work}

\begin{figure*}[t!]
  \centering
  \includegraphics[width=\textwidth]{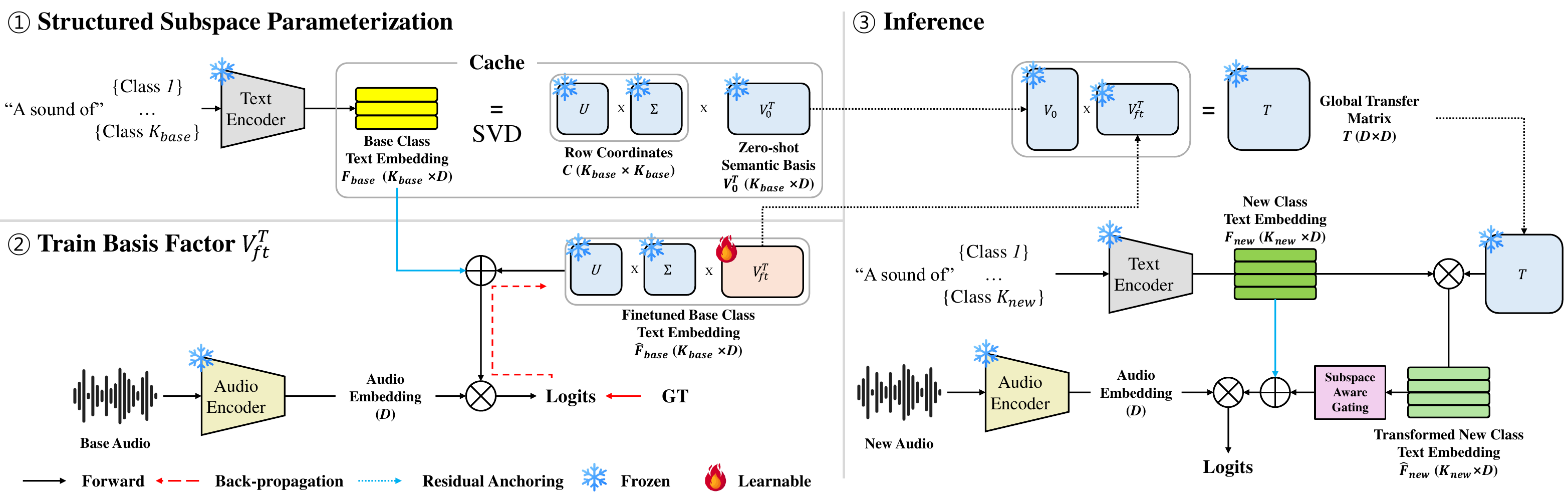}  
    \caption{
    Overview of Subspace Tuning (\textbf{SubT}). Starting from zero-shot base-class text embeddings, we compute $F_{\mathrm{base}} = U\Sigma V_0^\top$, keep the class-coordinate factor $C=U\Sigma$ fixed, and optimize the shared factor $V_{\mathrm{ft}}^\top$ for few-shot adaptation with residual anchoring to the zero-shot prototypes. New-class embeddings are not optimized during training. At inference, the learned adaptation is transferred through the base semantic subspace using the class-agnostic mapping $T=V_0V_{\mathrm{ft}}^\top$, and Subspace-aware Gating modulates the transferred component according to base-subspace alignment.
    }
  \label{fig:overview}
\end{figure*}

\paragraph{Few-Shot Adaptation for Audio-Language Models.}
Several studies have explored the adaptation of audio-language models (ALMs) under limited supervision. 
For example, test-time adaptation methods have been proposed for ALMs~\citep{test_time_adaptation, emo-tta}.
CLEP-DG~\citep{clep_dg} similarly uses text-guided prompt optimization to model acoustic context, particularly for speech emotion recognition.
PALM~\citep{palm} further addresses few-shot adaptation by directly tuning the embedding space.
However, although these methods improve downstream adaptation, the base-to-new generalization problem remains largely underexplored in ALMs.

\paragraph{Generalization under Few-Shot Tuning.}

Few-shot generalization has been studied much more extensively in vision--language models (VLMs). Prior work improves generalization by introducing additional inductive biases: CoCoOp~\citep{cocoop} uses instance-conditional prompts, KgCoOp~\citep{kgcoop} and LASP~\citep{lasp} regularize learned prompts toward handcrafted textual priors, and DePT~\citep{dept} decouples adaptation through auxiliary heads to separate task-specific fitting from transferable representations.
These methods are highly relevant to our work because they address the same few-shot generalization problem through modality-agnostic adaptation mechanisms on the prompt side.
More recently, SEPT~\citep{sept} addressed this gap in narrow-label audio settings by leveraging LLM-derived semantic neighbors to regularize the embedding-distance structure around each class.
In contrast, we perform geometry-aware adaptation directly in the output embedding space, yielding a simple and effective method for few-shot generalization in narrow-label audio--language tasks.
SubT is also conceptually related to subspace-preserving adaptation methods~\citep{franke2024preserving, wang2026complementary, lu2025controlled}, which constrain weight-space updates to preserve pretrained knowledge.


\section{Methodology}
\label{sec:method}

\subsection{Preliminaries}
\label{sec:preliminaries}

Audio--Language Models (ALMs) learn aligned audio and text representations via contrastive learning on large-scale paired data, enabling cross-modal retrieval and zero-shot recognition. In zero-shot audio classification, we use the pretrained ALM directly without task-specific fine-tuning.
Given an input audio sample, let $\mathbf{x} \in \mathbb{R}^{1 \times D}$ denote its $\ell_2$-normalized audio embedding extracted from the frozen audio encoder.
For a $K$-way classification task with class names $\{c_i\}_{i=1}^{K}$, we construct class-descriptive prompts (e.g., ``This is a sound of \{class\}'') and obtain the corresponding $\ell_2$-normalized text embeddings.
Stacking them row-wise yields a matrix $F \in \mathbb{R}^{K \times D}$, where the $i$-th row $\mathbf{f}_i \in \mathbb{R}^{1 \times D}$ represents the $i$-th class prototype in the shared embedding space.
The zero-shot prediction probability is then defined as
\begin{equation}
\label{eq:pred_zs_audio}
p(y \mid \mathbf{x}) =
\frac{\exp\!\big(\mathbf{x}\mathbf{f}_y^\top / \tau\big)}
{\sum_{i=1}^{K}\exp\!\big(\mathbf{x}\mathbf{f}_i^\top / \tau\big)},
\end{equation}
where $\tau$ is a temperature parameter.

\subsection{Subspace Tuning for Few-Shot Adaptation}
\label{sec:subt}

\paragraph{Setup and Motivation.}
In the few-shot adaptation setup, we are given a small labeled dataset from a set of \emph{base} classes of size $K_{\text{base}}$, and adapt only the class text embeddings in the embedding space.
Let $F_{\text{base}} \in \mathbb{R}^{K_{\text{base}} \times D}$ denote the matrix of zero-shot base-class text embeddings.
As discussed in Sec.~\ref{sec:intro}, unconstrained adaptation can induce substantial zero-shot drift, altering inter-class relations and moving the adapted prototypes away from their pretrained references.
Our goal is therefore to obtain a task-adaptive update while retaining a stable connection to the pretrained semantic structure.

As illustrated in Fig.~\ref{fig:overview}, SubT combines a geometry-aware SVD parameterization with residual anchoring.
The former parameterizes base-class adaptation through the zero-shot class geometry, while the latter retains an explicit connection to the original zero-shot prototypes.
The resulting factorization further provides a shared, subspace-mediated transformation that can be transferred to unseen classes at inference time.

\paragraph{Structured Subspace Parameterization.}
To parameterize adaptation according to the geometry of the zero-shot base-class prototypes, we apply Singular Value Decomposition (SVD) to $F_{\text{base}}$:
\begin{equation}
\label{eq:svd}
F_{\text{base}} = U\Sigma V_0^\top,
\end{equation}
where $U \in \mathbb{R}^{K_{\text{base}} \times r}$, $\Sigma \in \mathbb{R}^{r \times r}$, and $V_0 \in \mathbb{R}^{D \times r}$.
We use the economy (thin) SVD with $r=\operatorname{rank}(F_{\text{base}})$, computed separately for each base-class split; in our ALM setup, this typically gives $r=K_{\text{base}}$ rather than the total number of dataset classes, since $K_{\text{base}} \ll D$.

Under this decomposition, the matrix
\begin{equation}
C \triangleq U\Sigma \in \mathbb{R}^{K_{\text{base}} \times r}
\label{eq:rank}
\end{equation}
contains the coordinates of the base-class prototypes in the $r$-dimensional row space of $F_{\text{base}}$, while the columns of $V_0$ form an orthonormal basis for this subspace.
We freeze $C$ and optimize only the basis factor
$V_{ft}^\top \in \mathbb{R}^{r \times D}$, initialized from $V_0^\top$.
The resulting basis-transformed base-class prototypes are
\begin{equation}
\label{eq:factor_update}
\widehat{F}_{\text{base}}
=
\operatorname{Norm}\!\left(CV_{ft}^\top\right),
\end{equation}
where $\operatorname{Norm}(\cdot)$ denotes row-wise $\ell_2$-normalization.
When $F_{\text{base}}$ has full row rank, as in our default setting where
$r=K_{\text{base}}$, $C$ is square and invertible.
Therefore, this parameterization does not reduce the set of attainable
base-class prototype matrices relative to direct per-class optimization
and retains the same $K_{\text{base}}\times D$ parameter count.
Its role in this regime is instead to alter the optimization geometry
while providing a shared factorization that can be transferred to unseen classes.

\paragraph{Residual Anchoring.}
The SVD parameterization alone does not prevent the learned prototypes from moving substantially away from their zero-shot references.
Under scarce base-class supervision, the optimized representation can therefore still become overly specialized to the seen classes.
To stabilize adaptation, we retain an explicit zero-shot component through
residual anchoring:
\begin{equation}
\label{eq:residual}
F_{\text{base}}^{\text{tuned}} = \operatorname{Norm}\big(F_{\text{base}} + \widehat{F}_{\text{base}}\big).
\end{equation}
This fixed residual connection keeps the tuned prototype explicitly anchored to
its zero-shot counterpart while still allowing task-specific adjustment.
In this view, the SVD parameterization provides a geometry-aware
reparameterization of base-class adaptation, while residual anchoring
directly controls displacement from the zero-shot reference.

\paragraph{Training Objective.}
Given a few-shot labeled dataset
$\mathcal{D}=\{(\mathbf{x}_n, y_n)\}_{n=1}^{N}$
from the base classes, we optimize $V_{ft}^\top$ by minimizing the cross-entropy loss using the tuned base-class embeddings $F_{\text{base}}^{\text{tuned}}$:
\begin{equation}
\label{eq:ce}
\mathcal{L}_{\text{CE}}
=
-\frac{1}{N}\sum_{n=1}^{N}
\log p\big(y_n \mid \mathbf{x}_n; F_{\text{base}}^{\text{tuned}}\big).
\end{equation}
During training, the prediction space is restricted to the $K_{\text{base}}$ seen classes, and unseen classes are incorporated only at inference time through basis transfer.
Crucially, all ALM parameters remain frozen. Since $V_{ft}^\top$ is optimized entirely in the embedding space, SubT does not require backpropagation through the text encoder.

\subsection{Zero-shot Inference on Unseen Classes}

At test time, we must transfer the learned adaptation to a disjoint set of \emph{new} classes of size $K_{\text{new}}$, without using any new-class supervision.
The new-class embeddings start from their zero-shot prototypes and are adapted only through the shared transformation mediated by the base semantic subspace.
A key challenge is negative transfer: blindly applying a base-optimized update to semantically distant classes can distort otherwise robust zero-shot prototypes.
To address this, we first transfer the learned adaptation through the base semantic subspace to the new classes and then modulate its strength according to base-subspace alignment.

\paragraph{Shared Basis-Shift Transfer.}
Let $F_{\text{new}} \in \mathbb{R}^{K_{\text{new}} \times D}$ denote the matrix of zero-shot text embeddings for the new classes.
Because $C$ is defined only for base classes, unseen classes are handled by transferring only the learned shared basis shift.
Specifically, we construct a class-agnostic transfer operator using the zero-shot base basis $V_0$ and the learned shared factor $V_{ft}^\top$:
\begin{equation}
\label{eq:T}
T \triangleq V_0 V_{ft}^\top \in \mathbb{R}^{D \times D}.
\end{equation}
Here, $T \in \mathbb{R}^{D \times D}$, so it can be directly applied to $F_{\text{new}} \in \mathbb{R}^{K_{\text{new}} \times D}$.
Applying $T$ to $F_{\text{new}}$ yields a candidate basis-transferred embedding matrix:
\begin{equation}
\label{eq:new_raw}
\widehat{F}_{\text{new}} = \operatorname{Norm}(F_{\text{new}} T).
\end{equation}
This construction transfers the learned shared basis shift to unseen classes without introducing any new-class-specific parameters.

\paragraph{Subspace-Aware Gating.}
Not all unseen classes should receive the transferred update with equal strength.
If a new class is weakly aligned with the base semantic subspace, aggressively applying the transferred basis shift may cause negative transfer.
We therefore compute a {base-subspace alignment} score for each new-class embedding.

For the $i$-th new-class prototype $\mathbf{f}^{\text{new}}_i \in \mathbb{R}^{1 \times D}$, we measure the magnitude of its projection onto the frozen base basis $V_0$:
\begin{equation}
\label{eq:beta}
\beta_i \triangleq \left\| \mathbf{f}^{\text{new}}_i V_0 \right\|_2.
\end{equation}
Since $\mathbf{f}^{\text{new}}_i$ is $\ell_2$-normalized and $V_0$ has orthonormal columns, $\beta_i \in [0,1]$ in theory.
A larger $\beta_i$ indicates that the new class is better aligned with the base subspace, while a smaller $\beta_i$ suggests weaker alignment.

We then combine the original zero-shot prototype with the basis-transferred candidate through a gated residual connection:
\begin{equation}
\label{eq:new_final}
\mathbf{f}^{\text{new, final}}_i
=
\operatorname{Norm}\big(
\mathbf{f}^{\text{new}}_i
+
\beta_i \cdot \widehat{\mathbf{f}}^{\text{new}}_i
\big),
\end{equation}
where $\widehat{\mathbf{f}}^{\text{new}}_i$ is the $i$-th row of $\widehat{F}_{\text{new}}$.
This subspace-aware gating mechanism attenuates the transferred component for unseen classes with weak base-subspace alignment, thereby biasing the final prototype toward its original zero-shot representation.

We denote by \textbf{$\text{SubT}^{\dagger}$} the inference-time variant that applies subspace-aware gating when modulating the transferred component for unseen classes. The default \text{SubT} uses the same transfer without gating, which is equivalent to setting $\beta_i = 1$ for all unseen classes.


\begin{table*}[t!]
  \centering
  \footnotesize
  \scalebox{0.92}{
    \begin{tabular}{lcccccccccccc}
      \toprule
      \multicolumn{1}{l}{\multirow{2}{*}{\makecell[c]{Method}}} 
      & \multicolumn{3}{c}{Avg. over 11 datasets} 
      & \multicolumn{3}{c}{Beijing-Opera} 
      & \multicolumn{3}{c}{NS-Instruments} 
      & \multicolumn{3}{c}{ESC50} \\
      \cmidrule(lr){2-4}\cmidrule(lr){5-7}\cmidrule(lr){8-10}\cmidrule(lr){11-13}
      & Base & New & H & Base & New & H & Base & New & H & Base & New & H \\
      \midrule
      Zero-shot & 62.66 & 61.17 & 60.02 & 79.74 & 51.20 & 62.20 & 46.24 & \textbf{67.39} & 54.84 & 64.90 & \textbf{66.40} & 65.51 \\
      CoOp~\textcolor{gray} & 84.49 & 52.36 & 62.62 & 97.78 & 51.47 & 67.31 & 64.46 & 60.58 & 61.85 & 96.10 & 55.80 & 70.47 \\
      CoCoOp~\textcolor{gray} & \underline{85.76} & 52.81 & 63.19 & \textbf{100.00} & \underline{52.53} & \underline{68.74} & \underline{67.61} & 55.75 & 58.99 & \underline{97.33} & 61.80 & 75.49 \\
      KgCoOp~\textcolor{gray} & 56.28 & 43.07 & 47.12 & 65.52 & 50.94 & 56.20 & 41.34 & 46.13 & 42.30 & 49.47 & 28.53 & 35.88 \\
      DePT~\textcolor{gray} & 83.09 & 55.94 & 64.41 & 86.87 & 50.48 & 61.79 & 66.09 & 65.18 & \textbf{65.46} & 96.83 & 58.67 & 72.89 \\
      SEPT~\textcolor{gray} & 81.95 & 55.34 & 63.95 & 89.87 & 46.94 & 60.71 & 65.72 & 65.37 & \underline{65.18} & 96.17 & \underline{65.40} & \textbf{77.77} \\
      CLIP-Adapter~\textcolor{gray} & 77.08 & 59.98 & 65.27 & \underline{99.68} & 51.20 & 67.65 & 54.22 & \underline{66.47} & 59.53 & 85.53 & 64.73 & 73.38 \\
      \rowcolor{cyan!10}
      \textbf{SubT} & \textbf{87.89} & \underline{62.49} & \underline{71.79} & \textbf{100.00} & \textbf{60.63} & \textbf{74.72} & \textbf{70.66} & 55.24 & 61.84 & \textbf{98.43} & 61.07 & 75.29 \\
      \rowcolor{cyan!10}
      \textbf{SubT\textsuperscript{\dag}} & \textbf{87.89} & \textbf{63.79} & \textbf{72.52} & \textbf{100.00} & 52.26 & 68.58 & \textbf{70.66} & 59.57 & 64.50 & \textbf{98.43} & 63.07 & \underline{76.77} \\
      \midrule

      \multicolumn{1}{l}{\multirow{2}{*}{\makecell[c]{Method}}} 
      & \multicolumn{3}{c}{ESC50-Actions} 
      & \multicolumn{3}{c}{UrbanSound8K} 
      & \multicolumn{3}{c}{CREMA-D} 
      & \multicolumn{3}{c}{RAVDESS} \\
      \cmidrule(lr){2-4}\cmidrule(lr){5-7}\cmidrule(lr){8-10}\cmidrule(lr){11-13}
      & Base & New & H & Base & New & H & Base & New & H & Base & New & H \\
      \midrule
      Zero-shot & 83.50 & 77.50 & 80.30 & 59.36 & 71.67 & 64.62 & 37.45 & \textbf{76.13} & 50.21 & 52.65 & 40.09 & 45.52 \\
      CoOp~\textcolor{gray} & 99.00 & 83.67 & 90.43 & 85.86 & 73.39 & 78.61 & \underline{57.05} & 28.55 & 34.39 & 63.89 & 35.10 & 45.30 \\
      CoCoOp~\textcolor{gray} & \underline{99.17} & 81.33 & 89.26 & \underline{87.88} & 70.94 & 78.10 & 54.56 & 18.43 & 25.36 & \underline{68.31} & 40.23 & 50.63 \\
      KgCoOp~\textcolor{gray} & 77.50 & 37.50 & 49.71 & 55.89 & 41.56 & 46.78 & 38.51 & 36.37 & 36.01 & 47.85 & 28.05 & 34.63 \\
      DePT~\textcolor{gray} & 98.67 & \underline{88.33} & \underline{93.15} & 86.01 & 72.96 & 78.54 & 50.94 & 39.09 & 39.18 & 61.87 & 29.37 & 39.41 \\
      SEPT~\textcolor{gray} & 98.50 & \textbf{90.67} & \textbf{94.38} & 85.36 & \underline{75.25} & 79.56 & 44.32 & 22.06 & 25.96 & 57.45 & 34.80 & 43.19 \\
      CLIP-Adapter~\textcolor{gray} & 96.33 & 70.17 & 80.37 & 79.07 & 64.40 & 70.50 & 47.57 & \underline{73.96} & 57.89 & 56.19 & 37.74 & 44.86 \\
      \rowcolor{cyan!10}
      \textbf{SubT} & \textbf{100.00} & 72.67 & 83.89 & \textbf{89.03} & 74.27 & \underline{80.70} & \textbf{60.30} & 71.17 & \underline{65.23} & \textbf{72.85} & \underline{42.29} & \underline{53.44} \\
      \rowcolor{cyan!10}
      \textbf{SubT\textsuperscript{\dag}} & \textbf{100.00} & 76.50 & 86.44 & \textbf{89.03} & \textbf{77.10} & \textbf{82.41} & \textbf{60.30} & 73.16 & \textbf{66.07} & \textbf{72.85} & \textbf{42.88} & \textbf{53.92} \\
      \midrule

      \multicolumn{1}{l}{\multirow{2}{*}{\makecell[c]{Method}}} 
      & \multicolumn{3}{c}{SESA} 
      & \multicolumn{3}{c}{GT-Music-Genre} 
      & \multicolumn{3}{c}{VocalSound} 
      & \multicolumn{3}{c}{TUT2017} \\
      \cmidrule(lr){2-4}\cmidrule(lr){5-7}\cmidrule(lr){8-10}\cmidrule(lr){11-13}
      & Base & New & H & Base & New & H & Base & New & H & Base & New & H \\
      \midrule
      Zero-shot & 82.35 & \textbf{97.30} & 89.20 & 68.32 & 48.15 & 56.47 & 67.17 & 52.19 & 58.74 & 47.58 & 24.83 & 32.57 \\
      CoOp~\textcolor{gray} & 93.63 & 73.87 & 81.92 & 89.44 & 33.67 & 48.47 & 92.24 & 47.01 & 62.25 & 89.98 & 32.81 & 47.87 \\
      CoCoOp~\textcolor{gray} & 93.14 & 86.49 & 89.47 & \underline{90.43} & 39.73 & 54.36 & \underline{93.72} & 41.65 & 57.52 & \underline{91.25} & 32.00 & 47.18 \\
      KgCoOp~\textcolor{gray} & 75.00 & 90.99 & 82.14 & 63.37 & 41.75 & 50.24 & 64.40 & 43.99 & 51.92 & 40.28 & 28.00 & 32.49 \\
      DePT~\textcolor{gray} & \textbf{96.57} & 93.69 & 95.05 & 88.45 & 42.42 & 57.23 & 91.98 & 42.98 & 58.56 & 89.71 & 32.15 & 47.21 \\
      SEPT~\textcolor{gray} & 93.14 & 79.28 & 84.95 & 89.77 & \underline{53.54} & 66.23 & 90.83 & 44.77 & 59.95 & 90.35 & 30.62 & 45.59 \\
      CLIP-Adapter~\textcolor{gray} & \underline{95.59} & \underline{96.40} & \underline{95.97} & 78.22 & \textbf{54.21} & 64.00 & 80.48 & 52.75 & 63.49 & 75.04 & 27.79 & 40.37 \\
      \rowcolor{cyan!10}
      \textbf{SubT} & \textbf{96.57} & 91.89 & 94.16 & \textbf{91.09} & 52.19 & \underline{66.33} & \textbf{94.21} & \underline{64.38} & \underline{76.47} & \textbf{93.69} & \underline{41.63} & \underline{57.62} \\
      \rowcolor{cyan!10}
      \textbf{SubT\textsuperscript{\dag}} & \textbf{96.57} & \underline{96.40} & \textbf{96.48} & \textbf{91.09} & \textbf{54.21} & \textbf{67.95} & \textbf{94.21} & \textbf{64.70} & \textbf{76.69} & \textbf{93.69} & \textbf{41.89} & \textbf{57.87} \\
      \bottomrule
    \end{tabular}
  }
  \caption{Comparison on base-to-new generalization across 11 audio benchmarks. We report accuracy on seen classes (Base), unseen classes (New), and their harmonic mean (H).}
  \label{tab:b2n_comparison}
\end{table*}

\section{Experiments}
\label{sec:experiments}

\subsection{Experimental Setup}

\paragraph{Datasets.}
Following the evaluation protocol of SEPT~\citep{sept}, we benchmark on a diverse suite of audio classification datasets spanning multiple acoustic domains and label semantics.
For instrument recognition, we use Beijing-Opera~\citep{beijing-opera} and NS-Instruments~\citep{ns-instruments}.
For sound event classification, we use ESC50~\citep{esc50}, ESC50-Actions, and UrbanSound8K~\citep{urbansound}.
For speech emotion recognition, we use CREMA-D~\citep{crema-d} and RAVDESS~\citep{ravdess}.
We further include SESA~\citep{sesa} for surveillance sound classification, TUT2017~\citep{tut2017} for acoustic scene classification, GT-Music-Genre~\citep{gt-music-genre} for music genre classification, and VocalSound~\citep{vocalsound} for vocal sound classification.

\paragraph{Backbone and Baselines.}
We adopt the audio and text encoders from Pengi~\citep{pengi} as the underlying audio--text backbone, following recent few-shot adaptation studies in ALMs, including PALM~\citep{palm} and SEPT.
We compare our Subspace Tuning against representative baselines from three groups:
(1) the zero-shot Pengi baseline~\citep{pengi};
(2) prompt-based adaptation methods, including CoOp~\citep{coop}, CoCoOp~\citep{cocoop}, KgCoOp~\citep{kgcoop}, DePT~\citep{dept}, and SEPT~\citep{sept};
and (3) embedding-space adaptation via CLIP-Adapter~\citep{clipadapter}.
We do not consider methods such as PALM, Tip-Adapter~\citep{tipadapter}, and ProKeR~\citep{proker}, since they are not directly compatible with base-to-new evaluation and do not provide adapted prototypes for unseen classes at test time.

\paragraph{Implementation Details and Evaluation Protocol.}
We follow the standard SEPT protocol.
For base-to-new evaluation, we split each dataset into disjoint base and new classes of equal size, train on the base classes, and report base accuracy, new accuracy, and their harmonic mean ($H$). For cross-dataset evaluation, we train on all source classes and directly evaluate on the target dataset without further adaptation, reporting source and target accuracies. Unless otherwise specified, we use 16 shots per training class and optimize only the learnable parameters while keeping both the audio and text encoders frozen. All results are averaged over three random seeds. Additional details are provided in Appendix~\ref{sec:appendix_details}.

\subsection{Experimental Results}

\paragraph{Base-to-New Generalization.}
Table~\ref{tab:b2n_comparison} reports base-to-new results on 11 audio benchmarks. Overall, \textbf{SubT\textsuperscript{\dag}} achieves the best average harmonic mean (H) of 72.52, outperforming the strongest prior baseline, CLIP-Adapter, by +7.25 points, while also attaining the highest average accuracy on unseen classes (63.79).
Even without gating, \textbf{SubT} improves the average H to 71.79, and Subspace-aware Gating further improves both the average new-class accuracy and harmonic mean, indicating that alignment-aware modulation is beneficial on average for in-domain unseen-class transfer.

A consistent pattern is that many prior adaptation methods improve base-class accuracy at the cost of new-class generalization. For example, CoOp, CoCoOp, DePT, and SEPT all improve average base accuracy over the zero-shot model but reduce average new-class accuracy. In contrast, both SubT variants improve over zero-shot in both average base accuracy and average new accuracy, yielding a substantially better base-to-new trade-off.
This advantage is also reflected on challenging datasets. On TUT2017 and RAVDESS, SubT\textsuperscript{\dag} achieves the best harmonic mean of 57.87 and 53.92, respectively, and it also attains the best overall H on CREMA-D. These results support the effectiveness of combining geometry-aware adaptation with zero-shot anchoring for improving transfer to unseen classes.

\paragraph{Cross-dataset Evaluation.}
Table~\ref{tab:cross_dataset_main} reports cross-dataset transfer, where adaptation is learned on a source dataset and directly evaluated on an unseen target dataset. This setting is more challenging because both the label space and the data distribution can shift across domains. Overall, the SubT variants achieve the strongest source accuracy in all three transfer settings, while target performance is more mixed under domain shift.

In Instrument Classification, \textbf{SubT} performs best overall, achieving the highest source and target accuracies. In Emotion Recognition, the zero-shot model remains strongest on the target dataset, highlighting the difficulty of maintaining target-side generalization for this source--target pair; nevertheless, \textbf{SubT} substantially improves source accuracy while achieving the best target accuracy among adapted methods. In Sound Event Classification, the SubT variants again achieve the strongest source accuracy, whereas SEPT and DePT obtain higher target accuracy, showing that stronger source-side adaptation does not necessarily translate into better target-side transfer under domain shift.

\begin{table}[t!]
  \centering
  \scalebox{0.57}{
  \begin{tabular}{lcccccc}
    \toprule
    \multirow{3}{*}{Method}
      & \multicolumn{2}{c}{Instrument Classif.}
      & \multicolumn{2}{c}{Emotion Recog.}
      & \multicolumn{2}{c}{Sound Event Classif.} \\
    \cmidrule(lr){2-3} \cmidrule(lr){4-5} \cmidrule(lr){6-7}
      & \makecell{NS-Inst.\\(Source)} & \makecell{Beijing.\\(Target)}
      & \makecell{RAV.\\(Source)} & \makecell{CREM.\\(Target)}
      & \makecell{ESC50-A.\\(Source)} & \makecell{UrbanS.\\(Target)} \\
    \midrule
    Zero-shot & 36.38 & 28.81 & 28.51 & \textbf{52.99} & 64.25 & 52.57 \\
    CoOp~\textcolor{gray} & 61.44 & 26.41 & 37.68 & 28.70 & 94.58 & 50.82 \\
    CoCoOp~\textcolor{gray} & \underline{66.53} & 34.74 & \underline{41.62} & 35.30 & \underline{97.42} & 55.18 \\
    KgCoOp~\textcolor{gray} & 20.01 & 31.50 & 11.27 & 36.09 & 18.33 & 21.03 \\
    DePT~\textcolor{gray} & 61.57 & 29.81 & 38.49 & 30.04 & 95.08 & \underline{56.25} \\
    SEPT~\textcolor{gray} & 59.93 & 27.54 & 36.59 & 19.39 & 94.58 & \textbf{59.34} \\
    CLIP-Adapter~\textcolor{gray} & 59.78 & 27.54 & 33.13 & 45.36 & 82.08 & 49.65 \\
    \rowcolor{cyan!10}
    \textbf{SubT} & \textbf{68.87} & \textbf{37.03} & \textbf{48.67} & \underline{49.27} & \textbf{98.42} & 48.72 \\
    \rowcolor{cyan!10}
    \textbf{SubT\textsuperscript{\dag}} & \textbf{68.87} & \underline{36.61} & \textbf{48.67} & 49.16 & \textbf{98.42} & 50.98 \\
    \bottomrule
  \end{tabular}
  }
  \caption{Comparison on cross-dataset evaluation. We report accuracy on the source dataset used for adaptation and the unseen target dataset.}
  \label{tab:cross_dataset_main}
\end{table}


These results indicate that SubT is particularly effective for source-side adaptation under cross-dataset transfer, while target-side generalization remains more sensitive to source-target label mismatch and domain shift; we analyze this behavior further in Appendix~\ref{sec:failure_case_analysis}.

\begin{figure}[t!]
  \centering
  \includegraphics[width=\linewidth]{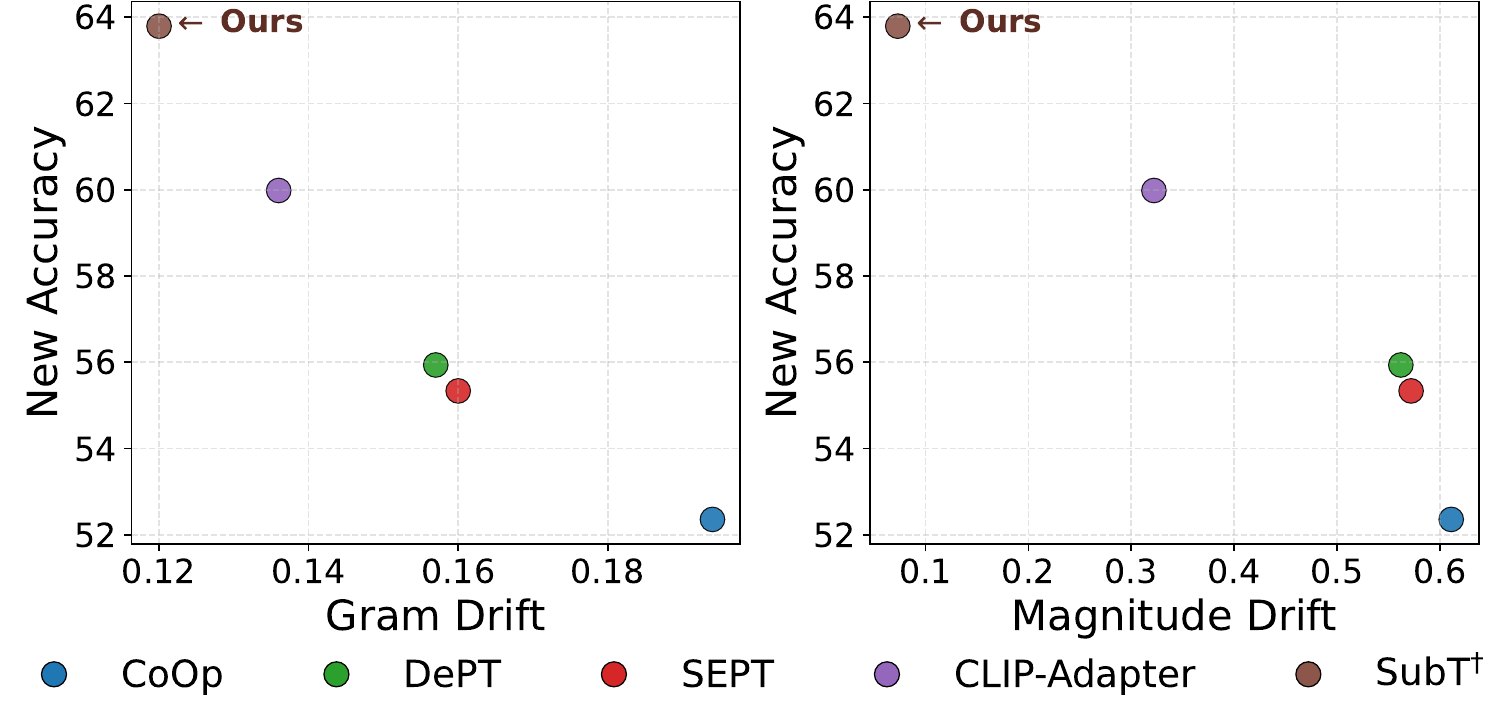}
  \caption{
    Relationship between new-class accuracy and two forms of drift, averaged over 11 datasets (left: Gram drift vs.\ new accuracy; right: magnitude drift vs.\ new accuracy).
    Overall, methods with smaller drift tend to achieve higher new-class accuracy, with our method showing the lowest drift and the strongest performance.
}
  \label{fig:drift_scatter}
\end{figure}

\begin{figure*}[t!]
  \centering
  \includegraphics[width=.9\textwidth]{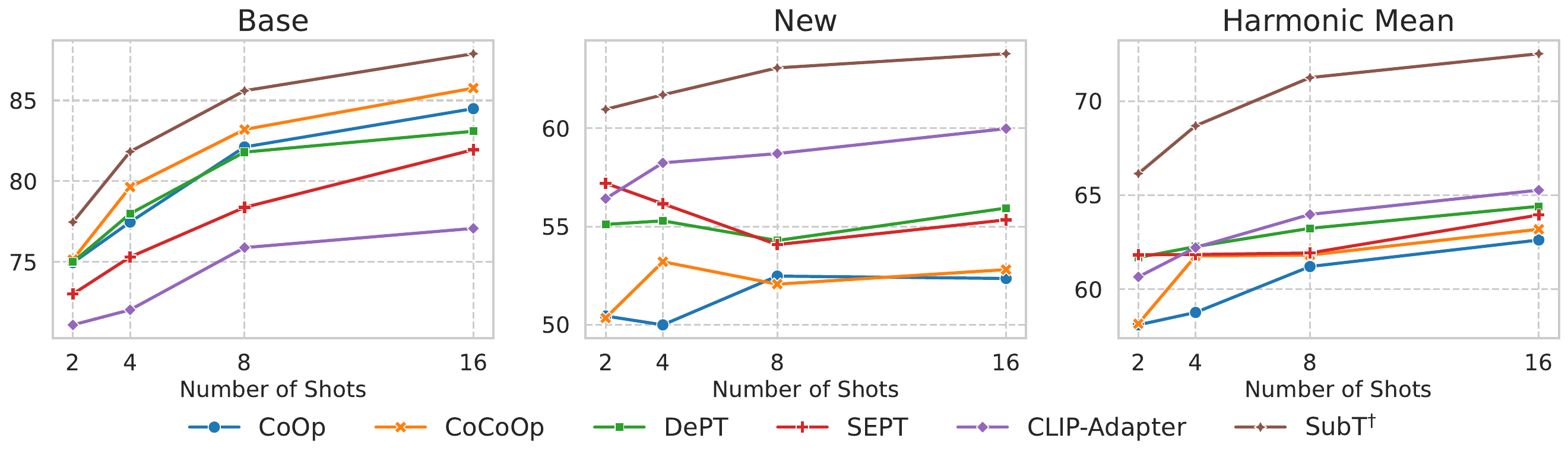}
  \caption{Performance dynamics across varying few-shot capacities (2, 4, 8, and 16 shots), averaged over 11 datasets. While baselines suffer from performance stagnation or degradation on new classes as shots increase, our SubT consistently scales across all metrics.}
  \label{fig:ablation_n_shots}
\end{figure*}

\subsection{Analysis}
\label{sec:analysis}

\paragraph{Geometry Drift and Accuracy.}
Figure~\ref{fig:drift_scatter} examines whether the two forms of zero-shot drift are empirically linked to unseen-class performance.
Averaged over 11 datasets, both plots show a negative association: methods with smaller Gram drift or magnitude drift tend to achieve higher new-class accuracy.
Our method lies in a favorable region of both plots, combining relatively low drift with strong new-class performance.
These observations are consistent with the hypothesis that maintaining a stable zero-shot reference may benefit unseen-class transfer, while not implying that minimizing drift alone is sufficient for generalization.

\paragraph{Computational Efficiency.}

Table~\ref{tab:efficiency_comparison} compares computational efficiency on TUT2017.
Among trainable adaptation methods, SubT\textsuperscript{\dag} offers a particularly strong accuracy--efficiency trade-off.
Under the same 8,192-parameter budget as CoOp, KgCoOp, SEPT, and CLIP-Adapter, SubT\textsuperscript{\dag} achieves the best harmonic mean of 57.87. It is also more training-efficient than most competing trainable methods, matching CLIP-Adapter and requiring less training time than standard prompt-tuning baselines by avoiding backpropagation through the text encoder. Inference is similarly lightweight, with latency nearly identical to the zero-shot baseline and other efficient adapters. Overall, these results show that SubT\textsuperscript{\dag} provides a strong accuracy--efficiency trade-off under a comparable model scale.

\begin{table}[t!]
  \centering
  \scalebox{0.7}{
  \begin{tabular}{lcccc}
    \toprule
    Method & \# Params & \makecell{Training \\ Time (s)} & \makecell{Inference \\ Time (ms)} & H \\
    \midrule
    Zero-shot            & 0        & 0     & 1.68 & 32.57 \\
    CoOp~\textcolor{gray}          & 8,192    & 23.53 & 1.68 & 47.87 \\
    CoCoOp~\textcolor{gray}        & 107,072  & 94.04 & 5.01 & 47.18 \\
    KgCoOp~\textcolor{gray}        & 8,192    & 24.96 & 1.68 & 32.49 \\
    DePT~\textcolor{gray}          & 28,680   & 24.75 & 1.68 & 47.21 \\
    SEPT~\textcolor{gray}         & 8,192    & 33.36 & 1.68 & 45.59 \\
    CLIP-Adapter~\textcolor{gray}  & 8,192    & 18.77 & 1.69 & 40.37 \\
    \rowcolor{cyan!10}
    \textbf{SubT\textsuperscript{\dag}}                    & 8,192    & 18.77 & 1.69 & 57.87 \\
    \bottomrule
  \end{tabular}
  }
  \caption{Efficiency comparison on the TUT2017 dataset.
  We report the number of learnable parameters, training time for 50 epochs, inference time per sample, and harmonic mean (H).}
  \label{tab:efficiency_comparison}
\end{table}

\subsection{Ablation Studies}
\label{sec:ablation_studies}

\begin{table}[t!]
  \centering
  \footnotesize
  \setlength{\tabcolsep}{4pt}
  \scalebox{0.75}{
  \begin{tabular}{c c c c c c c}
    \toprule
    Row &
    \makecell{Structured \\ Subspace \\ Parameterization} &
    \makecell{Residual \\ Anchoring} &
    \makecell{Subspace-aware \\ Gating} &
    Base & New & H \\
    \midrule
    1 &  &  &  & 87.41 & 42.95 & 56.07 \\
    2 & \checkmark &  &  & 87.64 & 43.67 & 57.67 \\
    3 &  & \checkmark &  & 87.68 & 60.10 & 69.47 \\
    4 & \checkmark & \checkmark &  & \textbf{87.89} & \underline{62.49} & \underline{71.79} \\
    \rowcolor{cyan!10}
    5 & \checkmark & \checkmark & \checkmark & \textbf{87.89} & \textbf{63.79} & \textbf{72.52} \\
    \bottomrule
  \end{tabular}
  }
  \caption{
    Ablation on the core components of SubT\textsuperscript{\dag} for base-to-new generalization: Structured Subspace Parameterization, Residual Anchoring, and Subspace-aware Gating.
    Results are averaged over 11 datasets.
  }
  \label{tab:ablation_component}
\end{table}

\paragraph{Component Ablation.}
Table~\ref{tab:ablation_component} ablates the three components of SubT for base-to-new generalization.
Row~1 directly optimizes free per-class base prototypes from the zero-shot initialization and transfers the learned adaptation to unseen classes through a global linear transform.

Structured Subspace Parameterization (SSP) replaces direct prototype optimization with the SVD-coordinate parameterization, fixing $U\Sigma$ and optimizing the shared factor.
Under the default full-rank setting, SSP does not reduce the set of attainable base-class prototypes, but it yields a modest improvement over direct per-class optimization, consistent with the different optimization geometry induced by the parameterization.
Residual Anchoring (RA) provides a substantially larger improvement, indicating that retaining an explicit connection to the zero-shot prototypes is the primary stabilizing factor under scarce supervision.
Adding SSP on top of RA further improves the base-to-new trade-off, showing that the SVD parameterization provides a complementary benefit to zero-shot anchoring.
Finally, Subspace-aware Gating provides an additional inference-time improvement in average unseen-class performance.
Overall, the ablation identifies distinct contributions: RA provides the dominant stabilization against zero-shot drift, SSP provides a modest complementary gain through geometry-aware reparameterization, and gating further improves unseen-class transfer.

\paragraph{Impact of Shot Capacity.}
Figure~\ref{fig:ablation_n_shots} shows performance as the number of labeled examples per base class increases from 2 to 16.
As expected, all methods improve in base accuracy as shot capacity increases, since more labeled examples per base class allow the adaptation process to better fit the decision boundaries of the seen classes evaluated on the same distribution.
The more informative trend appears in new accuracy: for several baselines, increasing the number of base-class shots does not reliably improve unseen-class performance, and can even lead to stagnation or degradation at intermediate shot counts.
This suggests that additional base supervision can increase base-specific specialization without necessarily improving transferability.
In contrast, SubT exhibits a more consistent trend.
Both base and new accuracies improve more steadily as the shot capacity increases, yielding a stronger harmonic mean across all shot settings.
This behavior is consistent with SubT maintaining a more stable base-to-new trade-off as the amount of base-class supervision increases.


\section{Conclusion}
\label{sec:conclusion}

In this paper, we investigated the base-to-new trade-off in adapting pretrained Audio--Language Models (ALMs) through the lens of zero-shot drift in the text embedding space.
We introduced \textbf{Subspace Tuning (SubT)}, an embedding-space adaptation method that combines geometry-aware SVD reparameterization with residual anchoring and transfers the learned adaptation to unseen classes through the base semantic subspace.
Experiments across diverse audio benchmarks showed that SubT provides strong base-to-new generalization, competitive cross-dataset transfer, and low computational overhead.
These results highlight the benefit of retaining a stable zero-shot reference while transferring task adaptation through shared semantic structure for few-shot generalization in ALMs.

\begin{table}[t!]
  \centering
  \scalebox{0.85}{
  \begin{tabular}{lcccc}
    \toprule
    Method & Base & New & H & \# params \\
    \midrule

    \multicolumn{1}{l}{Zero-shot} & 72.43 & 68.14 & 70.22 & 0 \\
    \multicolumn{1}{l}{CoOp}     & \textbf{76.65} & \textbf{69.65} & \textbf{72.98} & 8,192 \\
    \multicolumn{1}{l}{CLIP-Adapter} & 	74.53 & 68.81 & 71.55 & 8,192 \\
    \rowcolor{cyan!10}
    \multicolumn{1}{l}{SubT\textsuperscript{\dag}}     & \underline{76.47} & \underline{69.03} & \underline{72.56} & 256,000 \\

    \bottomrule
  \end{tabular}
  }
    \caption{
    Comparison on ImageNet.
    SubT improves over zero-shot and CLIP-Adapter, showing that the proposed embedding-space adaptation can also be applied beyond audio; CoOp remains slightly higher in this dense-label setting.
}

  \label{tab:ours_clip}
\end{table}

\section{Discussion}
\paragraph{Discussion.}
\label{sec:ablation_clip_main}
Although we focus on ALMs, SubT is not inherently audio-specific: it operates on class text embeddings in a shared multimodal space.
To examine this broader applicability, we apply SubT to CLIP~\cite{clip} on ImageNet~\citep{imagenet}.
As shown in Table~\ref{tab:ours_clip}, SubT improves over both the zero-shot baseline and CLIP-Adapter, showing that the proposed embedding-space adaptation can also be applied beyond audio.
While CoOp shows weaker generalization in narrow-label ALM benchmarks, it is slightly stronger in this dense 500-base-class regime, where broader label coverage may itself provide stronger implicit regularization during adaptation.
This result clarifies the scope of SubT: it is a modality-agnostic embedding-space method whose clearest advantage appears in narrow-label settings, where limited base-class supervision provides less semantic coverage and maintaining a stable zero-shot reference becomes particularly important.
Detailed analysis is provided in Sec.~\ref{sec:ablation_clip}.


\section{Limitations}
\label{sec:limitations}

While SubT demonstrates strong generalization, we note several limitations.
First, because both the SVD parameterization and unseen-class transfer are constructed from pretrained zero-shot text embeddings, the effectiveness of SubT depends on the quality and task relevance of the underlying semantic representation.
In highly specialized domains or when the unseen label space is weakly aligned with the base semantic subspace, the learned adaptation may transfer less reliably, limiting the benefit over the original zero-shot model.
Second, the number of learnable parameters scales linearly with the number of base classes.
As a result, SubT is most efficient in narrow-label audio tasks, while its efficiency advantage diminishes in larger-label settings.
Developing adaptive low-rank or compressed parameterizations that reduce this cost without sacrificing transfer performance remains an important direction for future work.

\bibliography{custom}


\newpage
\clearpage

\appendix
\label{sec:appendix}
\section*{\centering\LARGE Appendix}
\startcontents[appendixtoc]
\printcontents[appendixtoc]{l}{1}{\setcounter{tocdepth}{2}}
\addtocontents{toc}{\protect\setcounter{tocdepth}{2}}
\definecolor{linkcolor}{HTML}{000000}
\newpage

\definecolor{linkcolor}{HTML}{ED1C24}

\section{Detailed Experimental Settings}
\label{sec:appendix_details}

\subsection{Datasets and Zero-shot Prompts}
The number of total and base classes for each dataset is summarized in Table~\ref{tab:datasets_and_prompts}, together with the dataset type and the hand-crafted prompt template used to construct class text embeddings.
We adopt these templates across all methods to compute zero-shot text embeddings and to initialize the base-class embedding matrix used for adaptation.

\subsection{Implementation Details}
\label{sec:impl_details}

For a fair comparison, we implement prompt-learning baselines within a unified CoOp-style framework.
In particular, KgCoOp, DePT, and SEPT are all built on top of CoOp and follow the same prompt parameterization and optimization pipeline, differing only in their method-specific regularization or auxiliary components.
For the loss-balancing hyperparameters, we follow the original settings of the corresponding works, using $\lambda=8$ for KgCoOp and $\lambda=3$ for SEPT.
For CLIP-Adapter, we attach a lightweight two-layer MLP adapter with a hidden dimension of 4 to the text branch.
We choose this adapter size so that the number of learnable parameters is comparable to those of the other trainable baselines.
Following the pretrained Pengi backbone, the text embedding dimension is fixed to $D=1024$ throughout all experiments.
Accordingly, SubT learns $r \times D$ parameters in the embedding space; under the default full-rank setting used in our experiments, where $r=K_{\text{base}}$, this corresponds to $K_{\text{base}} \times 1024$ learnable parameters.
All experiments are conducted on a single RTX 4090 GPU.

\begin{table*}[ht!]
  \centering
  \small
  \scalebox{0.85}{
  \begin{tabular}{lccll}
    \toprule
    \textbf{Datasets} & \textbf{\# Total Classes} & \textbf{\# Base Classes} & \textbf{Type} & \textbf{Hand-crafted Prompt} \\
    \midrule
    Beijing-Opera & 4 & 2 & \multirow{2}{*}{Instrument Classification} & \multirow{2}{*}{``This is a sound of \{\} being played.''} \\
    NS-Instruments & 10 & 5 & & \\
    \midrule
    ESC50 & 50 & 25 & \multirow{3}{*}{Sound Event Classification} & \multirow{3}{*}{``This is a characteristic sound of \{\}.''} \\
    ESC50-Actions & 10 & 5 & & \\
    UrbanSound8K & 10 & 5 & & \\
    \midrule
    CREMA-D & 6 & 3 & \multirow{2}{*}{Emotion Recognition} & \multirow{2}{*}{``This is a human vocal sound of \{\} speech.''} \\
    RAVDESS & 8 & 4 & & \\
    \midrule
    SESA & 4 & 2 & Surveillance Sound Classification & ``This is a sound of \{\}.'' \\
    \midrule
    GT-Music-Genre & 10 & 5 & Music Analysis & ``This is a \{\} music.'' \\
    \midrule
    VocalSound & 6 & 3 & Vocal Sound Classification & ``This is a human vocal sound of \{\}.'' \\
    \midrule
    TUT2017 & 15 & 8 & Acoustic Scene Classification & ``This is a sound of \{\}.'' \\
    \bottomrule
  \end{tabular}
  }
  \caption{Summary of the evaluation datasets, their total and base-class counts, and corresponding hand-crafted prompt templates.}
  \label{tab:datasets_and_prompts}
\end{table*}

\subsection{Training Details}
\label{sec:training_details}

All learnable parameters were optimized for 50 epochs using SGD with momentum 0.9 and a learning rate of 0.0125.
We use a batch size of 16 for training and 256 for inference.
For all prompt-tuning baselines, we set the number of learnable context tokens to 16, where each context token is a 512-dimensional vector.
For all methods, the softmax temperature in Eq.~\eqref{eq:pred_zs_audio} is fixed to $\tau=0.01$.

\section{Additional Analysis}

\subsection{Comparison of Adaptation Methods}

Methods such as PALM~\citep{palm}, Tip-Adapter~\citep{tipadapter}, and ProKeR~\citep{proker} are not considered in our comparison because they are not compatible with the base-to-new evaluation protocol. PALM learns class-specific optimized text embeddings only for the training classes, while Tip-Adapter and ProKeR are training-free, cache-based methods that construct classifiers dynamically from support-set key-value pairs. These approaches are primarily designed for few-shot classification on seen classes and do not refine or provide transferable embeddings for unseen classes at test time. As a result, they cannot be applied to evaluating generalization on new classes, which is the central focus of our setting.

\begin{figure}[t!]
  \centering
  \includegraphics[width=\linewidth]{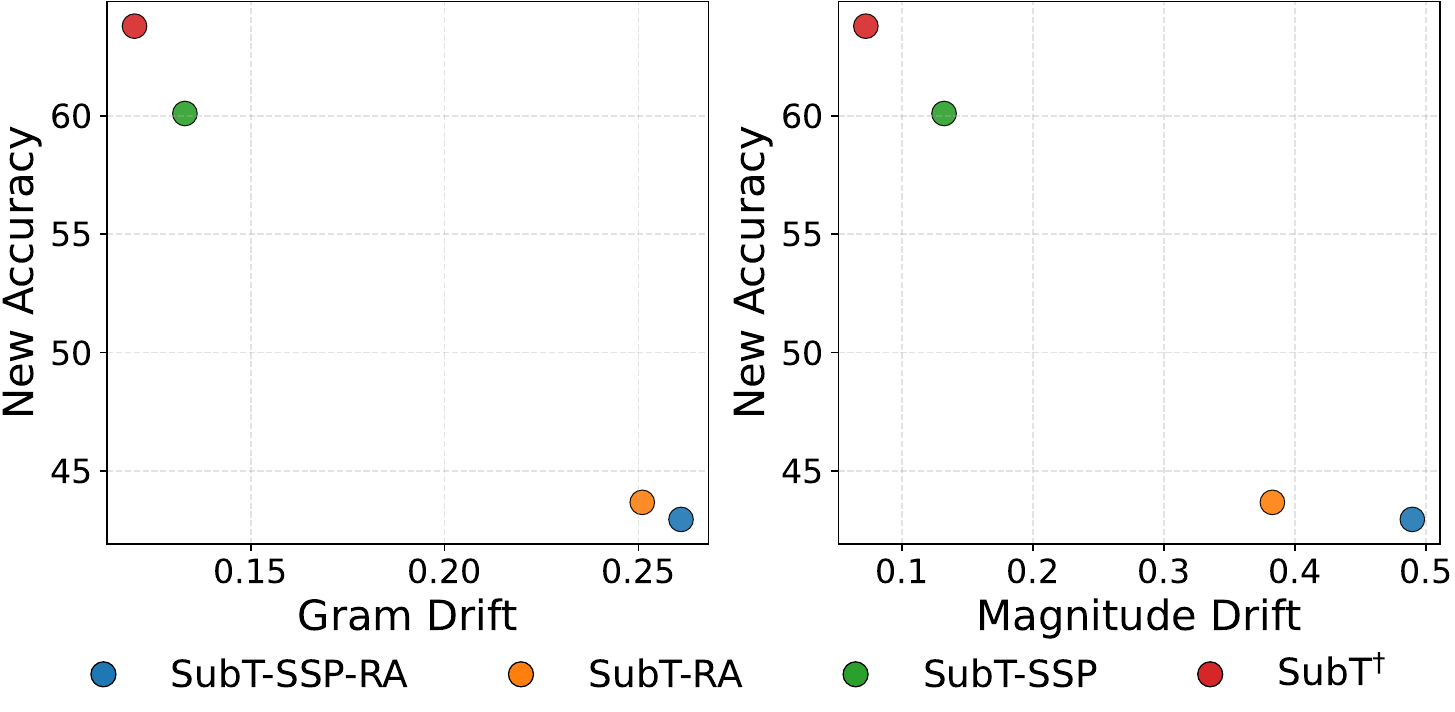}
  \caption{
    Relationship between new-class accuracy and two forms of drift for ablated variants of SubT, averaged over 11 datasets (left: Gram drift vs.\ new accuracy; right: magnitude drift vs.\ new accuracy).
    SSP denotes Structured Subspace Parameterization (Freeze $U\Sigma$), and RA denotes Residual Anchoring.
    Across the ablations, both Gram drift and magnitude drift show a negative association with new-class accuracy.
}
  \label{fig:drift_scatter_ablation}
\end{figure}

\subsection{Drift versus Performance in Component Ablations}

Figure~\ref{fig:drift_scatter_ablation} further examines how the core components of SubT affect unseen-class performance through two complementary forms of drift.
Across the evaluated ablations, lower Gram drift and magnitude drift are associated with higher new-class accuracy.

The two components show different empirical effects.
SSP alone yields a modest reduction in Gram drift relative to direct prototype optimization, whereas RA produces a substantially larger improvement in new-class performance together with reduced magnitude drift.
Combining SSP and RA gives the lowest observed drift and the strongest new-class accuracy among these ablations.

These results are consistent with complementary effects of geometry-aware reparameterization and zero-shot anchoring.
Importantly, under the default full-rank setting, the SSP gain should be interpreted as an effect of the parameterization rather than as evidence of a reduced base-class solution space.

\subsection{Effect of Simple Drift-Limiting Constraints}

The drift analyses above show that lower prototype drift is associated with stronger unseen-class accuracy, but this correlation alone does not fully isolate the role of the update constraint. We therefore evaluate a set of simple shared-transform baselines that directly restrict the form of the text-prototype update. All variants learn a single class-agnostic transform $T$ in the text-prototype space, apply it to zero-shot base-class prototypes during training, and transfer the same transform to zero-shot new-class prototypes at inference time. This protocol matches the base-to-new setting because it does not require new-class supervision and remains valid when base and new label sets are disjoint.

We compare three update families: (i) the unconstrained shared transform learns a full residual matrix $T=I+\Delta$, where $\Delta \in \mathbb{R}^{D \times D}$ is fully learnable; (ii) the low-rank shared transform constrains the residual update as $T=I+AB$, where $A \in \mathbb{R}^{D \times r}$ and $B \in \mathbb{R}^{r \times D}$, so the residual has rank at most $r$; and (iii) the hard-sparse shared transform again uses $T=I+\Delta$, but keeps only the top-$k$ entries of $\Delta$ active, yielding a sparse set of update directions. In all cases, transformed prototypes are row-wise normalized before classification.

\begin{table}[t!]
  \centering
  \small
  \scalebox{0.83}{
  \begin{tabular}{lccccc}
    \toprule
    Method & \makecell{Gram\\Drift} & \makecell{Mag.\\Drift} & Base & New & H \\
    \midrule
    Unconstrained & 0.152 & 0.184 & 87.64 & 61.53 & 70.95 \\
    Low-rank ($r=2$) & 0.070 & 0.151 & 85.17 & 62.87 & 70.47 \\
    Low-rank ($r=4$) & 0.047 & 0.116 & 87.46 & 63.14 & 71.58 \\
    Hard-sparse ($k=2048$) & 0.014 & 0.037 & 85.74 & 63.41 & 71.05 \\
    Hard-sparse ($k=4096$) & 0.016 & 0.044 & 86.04 & 63.46 & 71.23 \\
    Hard-sparse ($k=8192$) & 0.019 & 0.054 & 86.64 & 63.63 & 71.65 \\
    \midrule
    \rowcolor{cyan!10}
    \textbf{SubT\textsuperscript{\dag} } & 0.120 & 0.092 & \textbf{87.89} & \textbf{63.79} & \textbf{72.52} \\

    \bottomrule
  \end{tabular}
  }
  \caption{
    Simple shared-transform baselines averaged over 11 datasets.
    Constrained shared transforms reduce drift and improve unseen-class accuracy over the unconstrained baseline, while SubT\textsuperscript{\dag} achieves the best overall Base/New/H trade-off.
  }
  \label{tab:simple_drift_constraints}
\end{table}

Table~\ref{tab:simple_drift_constraints} supports the drift-control motivation through a direct intervention on the update family. Relative to the unconstrained shared transform, all constrained variants substantially reduce both Gram drift and magnitude drift, and all of them improve new-class accuracy. For example, the unconstrained transform obtains new accuracy of 61.53, whereas the constrained variants improve it to 62.87--63.63 while inducing much smaller drift.
Within these shared-transform baselines, the results show that explicitly restricting the update family can improve unseen-class accuracy relative to the unconstrained shared transform.

At the same time, the results show that minimizing drift as aggressively as possible is not sufficient. The hard-sparse variants achieve the smallest Gram drift, but they also sacrifice some base-class adaptation capacity and do not obtain the best harmonic mean. By contrast, SubT\textsuperscript{\dag} does not minimize Gram drift as strongly as the most restrictive sparse updates, yet it achieves the best Base, New, and H while maintaining substantially lower magnitude drift than the unconstrained baseline.
Overall, these results are consistent with a stability--plasticity trade-off: reducing excessive drift can benefit unseen-class transfer, but overly restrictive updates can sacrifice useful adaptation capacity.

\begin{figure}[t!]
  \centering
  \includegraphics[width=.95\linewidth]{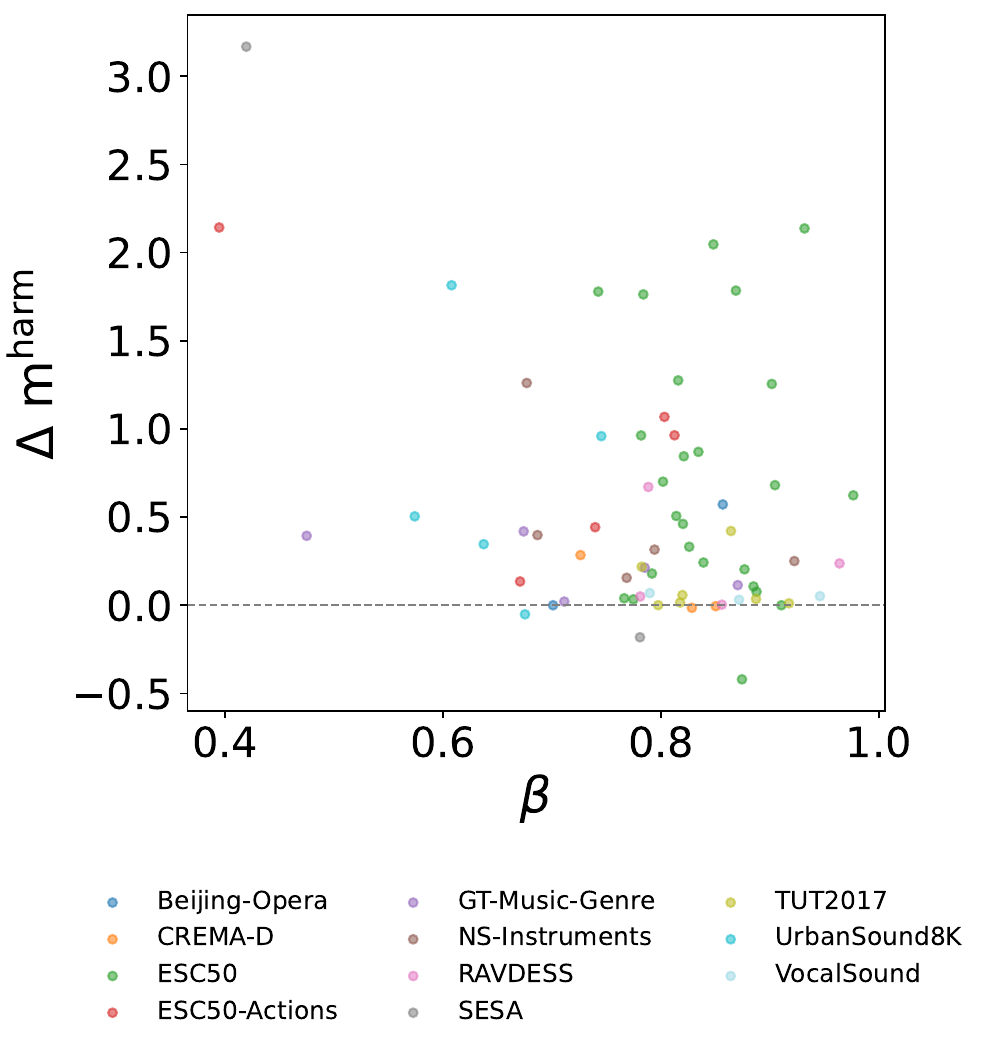}
  \caption{
    Relationship between class-level base-subspace alignment $\beta$ and the reduction in harmful transferred margin deficit, aggregated across all datasets.
    Each point corresponds to a new class.
    Most classes show positive $\Delta m_i^{\mathrm{harm}}$, indicating that subspace-aware gating generally reduces the harmful margin deficit associated with transferred adaptation on unseen classes.
  }
  \label{fig:beta_margin_all}
\end{figure}

\subsection{Beta and Harmful Update Reduction}
\label{sec:beta_margin}

To better understand the effect of subspace-aware gating, we analyze how the class-level alignment score $\beta$ relates to the reduction in harmful margin deficits on unseen classes.
For a test sample $x$ with ground-truth class $y$, we define the classification margin as
\begin{equation}
m(x) \triangleq s_y(x) - \max_{j \neq y} s_j(x),
\end{equation}
where $s_j(x)$ denotes the logit for class $j$.
To quantify the harmful effect of transferred adaptation relative to the zero-shot baseline, we define the positive margin deficit as
\begin{equation}
d_{\mathrm{harm}}(x; A) \triangleq \bigl(m_{\mathrm{ZS}}(x) - m_A(x)\bigr)_+,
\end{equation}
where $(\cdot)_+ \triangleq \max(\cdot, 0)$, $m_{\mathrm{ZS}}(x)$ is the margin under the zero-shot classifier, and $m_A(x)$ is the margin under method $A$.
This quantity measures the extent to which method $A$ yields a smaller margin than the zero-shot baseline, considering only the harmful part.
For each new class $i$, we then compute the class-aggregated reduction in harmful margin deficit as
\begin{equation}
\begin{split}
\Delta m_i^{\mathrm{harm}}
\triangleq\;
\mathbb{E}_{x:\,y=i} \Bigl[
d_{\mathrm{harm}}(x;\mathrm{SubT}) \\
- d_{\mathrm{harm}}(x;\mathrm{SubT}^{\dagger})
\Bigr],
\end{split}
\end{equation}
where $\mathrm{SubT}^{\dagger}$ denotes the variant with subspace-aware gating.
A positive $\Delta m_i^{\mathrm{harm}}$ indicates that gating reduces the harmful margin deficit introduced by transferred adaptation, thereby mitigating negative transfer relative to SubT.

Figure~\ref{fig:beta_margin_all} plots $\Delta m_i^{\mathrm{harm}}$ against the corresponding alignment score $\beta_i$, aggregated over all datasets.
Overall, most classes exhibit positive $\Delta m_i^{\mathrm{harm}}$, indicating that subspace-aware gating tends to reduce harmful effects of transferred adaptation on unseen classes.
At the same time, the dependence on $\beta_i$ is weak and not strictly monotonic.
This suggests that $\beta$ should not be interpreted as a strong predictor of the class-wise recovery achieved by gating.
Rather, $\beta$ serves as a lightweight heuristic control signal that modulates transfer conservatively according to the degree of base-subspace alignment.
In this sense, the role of $\beta$ is not to precisely predict the amount of harm reduction, but to bias adaptation toward safer transfer on unseen classes.
The actual benefit of gating remains class- and dataset-dependent, which is consistent with the broad spread of points in Fig.~\ref{fig:beta_margin_all}.

This class- and dataset-dependent behavior also helps explain the Beijing-Opera case in Table~\ref{tab:b2n_comparison}, where SubT\textsuperscript{\dag} yields lower new-class accuracy than ungated SubT. Beijing-Opera contains only four total classes, so the base-to-new split induces an extremely small base subspace. In such a low-dimensional regime, the alignment score $\beta$ can become less stable and may over-suppress a transferred update that is beneficial for some unseen classes. Nevertheless, ungated SubT remains strong on this dataset, improving new-class accuracy from 51.20 to 60.63 and H from 62.20 to 74.72 over zero-shot. Thus, subspace-aware gating should be interpreted as a lightweight conservative mechanism that improves average unseen-class performance across datasets, while its effect can vary across individual label splits.

\begin{table}[t!]
  \centering
  \small
  \scalebox{0.82}{
  \begin{tabular}{lcccc}
    \toprule
    Source $\rightarrow$ Target
    & $\beta \uparrow$
    & $D_{\mathrm{NN}} \downarrow$
    & $D_{\mathrm{OT}} \downarrow$
    & \makecell{Target Gain\\vs. ZS $\uparrow$} \\
    \midrule
    NS-Inst. $\rightarrow$ Beijing
    & 0.7920 & 0.3943 & 0.4697 & +7.80 \\
    ESC50-A. $\rightarrow$ UrbanS.
    & 0.6036 & 0.6058 & 0.6592 & -1.59 \\
    \bottomrule
  \end{tabular}
  }
  \caption{
    Source-target label-space compatibility in the frozen text embedding space.
    Target gain is measured as the target accuracy difference between SubT\textsuperscript{\dag} and the zero-shot baseline.
  }
  \label{tab:failure_case_compatibility}
\end{table}

\subsection{Analysis of Cross-Dataset Target Transfer}
\label{sec:failure_case_analysis}

Table~\ref{tab:cross_dataset_main} shows that SubT improves source-side adaptation in the cross-dataset setting, but does not always improve target-side transfer. The most informative target-side exception is ESC50-Actions $\rightarrow$ UrbanSound8K, where both datasets belong to sound event classification, yet SubT\textsuperscript{\dag} underperforms the zero-shot model on the target dataset. This suggests that the coarse task category alone is insufficient to determine whether a source-learned basis shift will transfer reliably.

To analyze this behavior, we measure source-target label-space compatibility in the frozen text embedding space. This is directly relevant to SubT because cross-dataset transfer applies a basis shift learned from source-class text prototypes to target-class text prototypes. Let $F_S \in \mathbb{R}^{K_S \times D}$ and $F_T \in \mathbb{R}^{K_T \times D}$ denote the row-normalized source and target class text embedding matrices, respectively. We consider three complementary metrics. First, we compute the average target projection norm onto the source semantic basis,
\begin{equation}
\beta_{S \rightarrow T}
= \frac{1}{K_T} \sum_{i=1}^{K_T}
\left\| f_i^{(T)} V_S \right\|_2,
\end{equation}
where $V_S$ is obtained from the SVD of $F_S$. Larger $\beta_{S \rightarrow T}$ indicates stronger alignment between the target label space and the source subspace. Second, we compute a symmetric nearest-neighbor cosine distance,
\begin{equation}
\begin{split}
D_{\mathrm{NN}} = 1 - \frac{1}{2} \Bigg[
&\frac{1}{K_T} \sum_{i=1}^{K_T}
\max_j \cos\!\left(f_i^{(T)}, f_j^{(S)}\right) \\
&+ \frac{1}{K_S} \sum_{j=1}^{K_S}
\max_i \cos\!\left(f_j^{(S)}, f_i^{(T)}\right)
\Bigg],
\end{split}
\end{equation}
where smaller values indicate stronger local label-space similarity. Third, we compute an optimal-transport distance with cosine cost,
\begin{equation}
D_{\mathrm{OT}} = \sum_{i,j} \pi^*_{ij} C_{ij},
\quad
C_{ij}=1-\cos\!\left(f_i^{(T)}, f_j^{(S)}\right),
\end{equation}
where $\pi^*$ is the optimal transport plan between uniform source and target class distributions. Smaller $D_{\mathrm{OT}}$ indicates stronger global compatibility between the two label distributions.

\begin{table*}[t!]
  \centering
  \small
  \setlength{\tabcolsep}{4pt}
  \begin{tabular}{llccccc}
    \toprule
    Dataset & Method & \# Params & \makecell{Train\\Time (s)} & Base & New & H \\
    \midrule
    \multirow{4}{*}{\makecell{Beijing-Opera\\($K_{\mathrm{base}}=2$)}}
    & CoOp & 8,192 & 51 & 97.78 & \underline{51.47} & 67.31 \\
    & CLIP-Adapter (default) & 8,192 & 47 & \underline{99.68} & 51.20 & \underline{67.65} \\
    & CLIP-Adapter (matched) & 2,048 & 43 & 85.56 & 51.20 & 63.72 \\
    \rowcolor{cyan!10} \cellcolor{white}
    & \textbf{SubT\textsuperscript{\dag}} & {2,048} & {43} & \textbf{100.00} & \textbf{52.26} & \textbf{68.58} \\
    \midrule
    \multirow{4}{*}{\makecell{ESC50\\($K_{\mathrm{base}}=25$)}}
    & CoOp & 8,192 & 154 & 96.10 & 55.80 & 70.47 \\
    & CLIP-Adapter (default) & 8,192 & 118 & 85.53 & \textbf{64.73} & \underline{73.38} \\
    & CLIP-Adapter (matched) & 24,576 & 121 & \underline{98.27} & 58.07 & 72.92 \\
    \rowcolor{cyan!10} \cellcolor{white}
    & \textbf{SubT\textsuperscript{\dag}} & 25,600 & 121 & \textbf{98.43} & \underline{63.07} & \textbf{76.77} \\
    \midrule
    \multirow{4}{*}{\makecell{ImageNet\\($K_{\mathrm{base}}=500$)}}
    & CoOp & 8,192 & 503 & \textbf{76.65} & \textbf{69.65} & \textbf{72.98} \\
    & CLIP-Adapter (default) & 8,192 & 118 & 74.53 & 68.81 & 71.55 \\
    & CLIP-Adapter (matched) & 262,144 & 125 & 75.78 & 68.82 & 72.14 \\
    \rowcolor{cyan!10} \cellcolor{white}
    & \textbf{SubT\textsuperscript{\dag}} & {256,000} & {125} & \underline{76.47} & \underline{69.03} & \underline{72.56} \\
    \bottomrule
  \end{tabular}
  \caption{
    Parameter-matched comparison with CLIP-Adapter across label-space scales.
    The matched CLIP-Adapter adjusts its hidden dimension to use a parameter budget close to SubT\textsuperscript{\dag}.
  }
  \label{tab:param_matched_clip_adapter}
\end{table*}

Table~\ref{tab:failure_case_compatibility} provides quantitative support for the failure-case interpretation. The successful NS-Instruments $\rightarrow$ Beijing-Opera pair has stronger source-target compatibility, with a higher subspace alignment score ($\beta=0.7920$) and smaller local and global distances ($D_{\mathrm{NN}}=0.3943$, $D_{\mathrm{OT}}=0.4697$). Consistently, this pair yields a positive target-side gain over zero-shot. In contrast, ESC50-Actions $\rightarrow$ UrbanSound8K has substantially lower alignment ($\beta=0.6036$) and larger text-space distances ($D_{\mathrm{NN}}=0.6058$, $D_{\mathrm{OT}}=0.6592$), matching the observed target-side degradation.

This analysis explains why sharing the same coarse task category is not sufficient for successful cross-dataset transfer. ESC50-Actions is dominated by human action- and body-related sounds, whereas UrbanSound8K contains urban ambient and machine-related events. Under this semantic mismatch, the basis shift learned from ESC50-Actions can become overly specialized to the source label geometry and less compatible with UrbanSound8K.
These cases are consistent with the hypothesis that source--target compatibility in the pretrained text space influences the reliability of cross-dataset transfer; broader source--target evaluations would be needed to establish this relationship systematically.

\subsection{Application to Vision-Language Models (CLIP)}
\label{sec:ablation_clip}

The ImageNet experiment in Table~\ref{tab:ours_clip} examines whether the embedding-space formulation of SubT also transfers to a vision--language backbone.
SubT improves over the zero-shot baseline and CLIP-Adapter, but CoOp achieves a slightly higher harmonic mean in this dense-label setting.

One possible explanation is that the 500 base classes in ImageNet provide substantially broader semantic coverage than the narrow-label audio benchmarks, reducing the relative benefit of explicit zero-shot anchoring and subspace-mediated transfer.
However, the present experiment does not isolate label-space density as the causal factor.

Accordingly, the ImageNet result bounds rather than contradicts the scope of SubT:
the method remains competitive in a large-label VLM setting, while its clearest empirical advantage is observed in the narrow-label ALM regimes studied in the main experiments.

\subsection{Parameter-Matched Adapter Comparison Across Label Scales}
\label{sec:param_matched_clip_adapter}

We further compare SubT\textsuperscript{\dag} with CLIP-Adapter under matched parameter budgets across different label-space scales.
For each dataset, we adjust the hidden dimension of CLIP-Adapter so that its parameter count is close to that of SubT\textsuperscript{\dag}. We consider three label-space scales: Beijing-Opera ($K_{\mathrm{base}}=2$), ESC50 ($K_{\mathrm{base}}=25$), and ImageNet ($K_{\mathrm{base}}=500$).

Table~\ref{tab:param_matched_clip_adapter} shows that SubT\textsuperscript{\dag} outperforms the parameter-matched CLIP-Adapter in all three regimes, indicating that its improvement is not merely a consequence of larger model capacity. Instead, the shared-basis update provides a more effective output-space adaptation mechanism than increasing the size of a generic adapter. At the same time, the ImageNet result clarifies the scope of this advantage: CoOp remains slightly stronger and more parameter-efficient in the dense-label VLM setting, although it requires substantially longer training. Thus, SubT\textsuperscript{\dag} should be viewed as a text-encoder-backpropagation-free adaptation method that is especially attractive for narrow-label ALM tasks, while remaining competitive but not universally dominant in large-label regimes.

\section{Additional Experiments}
\label{sec:additional_experiments}

\begin{table}[t!]
  \centering
  \small 
  \begin{tabular}{lccc}
    \toprule
    Method & Base & New & H \\
    \midrule
    Zero-shot     & 79.58 & 68.37 & 70.99 \\
    CoOp         & 81.30 & 68.96 & 72.72 \\
    CoCoOp       & \underline{83.75} & \underline{71.95} & \underline{76.10} \\
    KgCoOp       & 83.45 & 68.12 & 73.02 \\
    DePT         & 80.75 & 66.48 & 70.64 \\
    SEPT         & 83.29 & 70.56 & 74.73 \\
    CLIP-Adapter & 80.68 & 66.05 & 70.15 \\
    \rowcolor{cyan!10}
    \textbf{SubT\textsuperscript{\dag} }         & \textbf{83.78} & \textbf{72.09} & \textbf{76.22} \\
    \bottomrule
  \end{tabular}
  \caption{Performance comparison on the MS-CLAP~\citep{clap23} backbone. Results are averaged over 10 audio benchmarks (excluding ESC50 due to GPU memory limits). Our SubT\textsuperscript{\dag} achieves the highest average harmonic mean among the compared methods while requiring zero inference overhead compared to heavy instance-conditional methods.}
  \label{tab:comparison_clap}
\end{table}

\subsection{Generalization to Alternative ALM Architectures}
\label{sec:ablation_clap}
To examine whether the effectiveness of SubT extends beyond the Pengi backbone,
we additionally evaluate it on MS-CLAP~\citep{clap23}.
As shown in Table~\ref{tab:comparison_clap}, SubT\textsuperscript{\dag} achieves the highest average harmonic mean among the evaluated methods (76.22), narrowly exceeding CoCoOp (76.10).
Unlike CoCoOp, SubT does not require an instance-conditional meta-network at inference time.
These results suggest that the proposed embedding-space adaptation can also be effective with another ALM architecture, although broader backbone evaluation would be required to establish architecture-independent generality.

\begin{table}[t!]
  \centering
  \small 
  \begin{tabular}{lccc}
    \toprule
    Anchoring Strategy &  Base & New & H \\
    \midrule
    Learnable Blending & \textbf{88.04} & 63.37 & 72.13 \\
    \rowcolor{cyan!10}
    Residual Anchoring & 87.89 & \textbf{63.79} & \textbf{72.52} \\
    \bottomrule
  \end{tabular}
  \caption{Ablation study on the anchoring mechanism. We compare our fixed 1:1 residual anchoring against a learnable blending strategy.
  The fixed residual yields a better Base-to-New trade-off than learnable scalar blending in this experiment.
  }
  \label{tab:ablation_anchoring}
\end{table}

\subsection{Residual vs. Learnable Anchoring}
\label{sec:ablation_anchoring}

In our main method, we formulate the final tuned embedding using a simple 1:1 residual anchoring: $F_{\text{base}}^{\text{tuned}} = \operatorname{Norm}(F_{\text{base}} + \widehat{F}_{\text{base}})$. A natural alternative is to introduce a learnable blending parameter $\alpha$ to dynamically balance the zero-shot prior and the task-specific update: $F_{\text{base}}^{\text{tuned}} = \operatorname{Norm}((1-\alpha) F_{\text{base}} + \alpha \widehat{F}_{\text{base}})$, where $\alpha$ is a learnable scalar initialized to 0.5. 

Table~\ref{tab:ablation_anchoring} compares the fixed 1:1 residual with a learnable blending coefficient.
Learnable blending yields slightly higher base accuracy, whereas the fixed residual achieves higher new-class accuracy and harmonic mean.
This result is consistent with retaining a fixed zero-shot contribution being beneficial for the base-to-new trade-off under few-shot supervision.
The fixed formulation also avoids introducing an additional learnable blending parameter.

\begin{table}[t!]
  \centering
  \small
  \begin{tabular}{lccc}
    \toprule
    $\lambda$ (Penalty Weight) & Base & New & H \\
    \midrule
    \rowcolor{cyan!10}
    $0$ & 87.89 & 63.79 & 72.52 \\
    $0.01$ & 87.04 & 64.42 & 72.66 \\
    $0.05$ & 87.61 & 64.17 & 72.69 \\
    $0.10$ & 88.06 & 63.92 & 72.56 \\
    $0.50$ & 87.88 & 63.78 & 72.49 \\
    $1$    & 88.08 & 63.93 & 72.59 \\
    $5$    & 86.52 & 63.32 & 71.62 \\
    $10$   & 86.94 & 63.09 & 71.51 \\
    $50$   & 63.01 & 53.77 & 55.14 \\
    \bottomrule
  \end{tabular}
  \caption{Ablation study on applying an orthogonality regularizer to the basis factor $V_{ft}^\top$.
  While weak regularization ($\lambda\le1$) yields only small performance variations, stronger regularization ($\lambda\ge5$) substantially degrades both base and new accuracy.
  }
  \label{tab:ablation_orthogonal}
\end{table}

\subsection{Effect of Orthogonality Regularization}
\label{sec:ablation_orthogonality}

SubT optimizes the shared factor $V_{ft}^\top$ without explicitly enforcing orthogonality.
A natural alternative is to encourage $V_{ft}^\top$ to remain row-orthonormal, which would more strongly preserve the pairwise geometry of the zero-shot base prototypes.
Specifically, if $V_{ft}^\top V_{ft}=I$, the adapted base prototypes preserve the zero-shot Gram matrix:
\begin{equation}
\begin{aligned}
(CV_{ft}^\top)(CV_{ft}^\top)^\top
&= C(V_{ft}^\top V_{ft})C^\top \\
&= CC^\top
= F_{\text{base}}F_{\text{base}}^\top .
\end{aligned}
\end{equation}
To examine whether encouraging this property benefits generalization, we augment the training objective with an orthogonality penalty:
\begin{equation}
\mathcal{L}_{\text{total}}
=
\mathcal{L}_{\text{CE}}
+
\lambda
\left\|
V_{ft}^\top V_{ft}-I
\right\|_F^2,
\end{equation}
where $\lambda$ controls the strength of the penalty and $I$ is the identity matrix.
Table~\ref{tab:ablation_orthogonal} reports the results for different values of $\lambda$.

Weak orthogonality penalties ($\lambda \leq 1$) produce only small variations in harmonic mean relative to the default setting, with the best result reaching 72.69 compared with 72.52 without the penalty.
In contrast, stronger penalties ($\lambda \geq 5$) substantially degrade performance, with the harmonic mean decreasing to 55.14 at $\lambda=50$.
These results suggest that additional orthogonality regularization is not necessary in the evaluated setting.
While stronger orthogonality encourages greater preservation of the zero-shot Gram structure, it also restricts scaling and other deformations that may be useful for task adaptation, thereby reducing adaptation flexibility.
Given the small differences under weak regularization, the degradation under strong regularization, and the additional hyperparameter $\lambda$, we retain the simpler formulation without explicit orthogonality regularization as our default.

\begin{table}[t!]
  \centering
  \small
  \begin{tabular}{lccc}
    \toprule
    Rank $r$ & Base & New & H \\
    \midrule
    2 & 59.71 & 25.67 & 35.88 \\
    4 & 88.47 & 24.75 & 38.57 \\
    6 & 92.83 & 40.66 & 56.49 \\
    \rowcolor{cyan!10}
    8 (default) & \textbf{93.69} & \textbf{41.89} & \textbf{57.87} \\
    \bottomrule
  \end{tabular}
  \caption{Ablation on the rank $r$ of the shared-basis parameterization on TUT2017, where $K_{\text{base}}=8$. The default full-rank setting gives the best overall performance.}
  \label{tab:rank_ablation}
\end{table}

\subsection{Effect of Rank $r$}
\label{sec:rank_ablation}

By default, SubT uses the full rank of the base-class embedding matrix. 
Accordingly, in Eq.~\eqref{eq:rank}, we set $r$ to the full rank of the split-specific base-class matrix $F_{\text{base}}$, which typically equals the number of base classes $K_{\text{base}}$ rather than the total number of classes.
To examine the sensitivity to this choice, we conduct an ablation study on TUT2017, where the base split has $K_{\text{base}}=8$, and vary the rank as $r \in \{2,4,6,8\}$.
The results are reported in Table~\ref{tab:rank_ablation}.

A clear trend emerges: aggressive rank truncation harms generalization, especially on unseen classes.
While $r=2$ substantially degrades both base and new performance, $r=4$ largely restores base accuracy but still yields poor transfer to new classes.
Increasing the rank to $6$ markedly improves performance, and the default full-rank setting $r=8$ achieves the best overall harmonic mean.
These results suggest that aggressive rank truncation removes capacity that is useful for both base-class adaptation and unseen-class transfer in this setting.
In contrast, using the full available rank yields the strongest overall trade-off on TUT2017, while also avoiding an additional rank-selection hyperparameter.

\begin{table}[t!]
\centering
\small
\caption{Licenses of datasets, pretrained models, and baseline implementations used in this work.}
\resizebox{\columnwidth}{!}{
\begin{tabular}{llcl}
\toprule
\textbf{Type} & \textbf{Asset} & \textbf{License} & \textbf{Source} \\
\midrule
Dataset & Beijing-Opera & MIT & \href{https://huggingface.co/datasets/MahiA/Beijing-Opera}{Hugging Face} \\
        & NS-Instruments & CC BY 4.0 & \href{https://huggingface.co/datasets/jg583/NSynth}{Hugging Face} \\
        & ESC50 & CC BY-NC 3.0 & \href{https://github.com/karolpiczak/ESC-50}{GitHub} \\
        & ESC50-Actions & CC BY-NC 3.0 & \href{https://github.com/karolpiczak/ESC-50}{GitHub} \\
        & UrbanSound8K & CC BY-NC 4.0 & \href{https://urbansounddataset.weebly.com/urbansound8k.html}{UrbanSound8K} \\
        & CREMA-D & ODbL 1.0 & \href{https://github.com/CheyneyComputerScience/CREMA-D}{GitHub} \\
        & RAVDESS & CC BY-SA 4.0 & \href{https://zenodo.org/records/1188976}{Zenodo} \\
        & SESA & CC BY 4.0 & \href{https://zenodo.org/records/3519845}{Zenodo} \\
        & GT-Music-Genre & MIT & \href{https://huggingface.co/datasets/MahiA/GT-Music-Genre}{Hugging Face} \\
        & VocalSound & CC BY-SA 4.0 & \href{https://github.com/YuanGongND/vocalsound}{GitHub} \\
        & TUT2017 & Non-commercial & \href{https://zenodo.org/records/400515}{Zenodo} \\
        & ImageNet & Non-commercial & \href{https://www.image-net.org/}{ImageNet} \\
\midrule
Model & Pengi & MIT & \href{https://github.com/microsoft/Pengi}{GitHub} \\
      & CLAP & MIT & \href{https://github.com/microsoft/CLAP}{GitHub} \\
      & CLIP & MIT & \href{https://github.com/OpenAI/CLIP}{GitHub} \\
\midrule
Baseline & CoOp & MIT & \href{https://github.com/kaiyangzhou/coop}{GitHub} \\
         & CoCoOp & MIT & \href{https://github.com/kaiyangzhou/coop}{GitHub} \\
         & KgCoOp & Unknown & \href{https://github.com/htyao89/KgCoOp}{GitHub} \\
         & DePT & GPL-2.0 & \href{https://github.com/Koorye/DePT/}{GitHub} \\
         & SEPT & Unknown & \href{https://github.com/jhyukjang/SEPT}{GitHub} \\
         & CLIP-Adapter & Unknown & \href{https://github.com/gaopengcuhk/CLIP-Adapter}{GitHub} \\
\bottomrule
\end{tabular}
}
\label{tab:licenses}
\end{table}

\section{AI Assistant Usage Statement}

During the preparation of this manuscript, we utilized Gemini 2.5 Pro strictly for linguistic refinement, including grammatical corrections and stylistic improvements. All scientific hypotheses, experimental analyses, and core intellectual contributions remain entirely the original work of the human authors.

\section{Licenses of Datasets and Models}
\label{sec:asset_licenses}

We summarize the licenses of all datasets, pretrained models, and baseline implementations used in this work in Table~\ref{tab:licenses}. 
All assets are used in accordance with their respective licenses.

\end{document}